\documentclass[pra,aps,12pt,showpacs,showkeys]{revtex4}
\usepackage{epsfig}

\begin{document}

\author{Li-Bin Fu }

\affiliation{Max-Planck-Institute for the Physics of Complex
systems, N\"{o}thnitzer Str. 38, 01187 Dresden, Germany, and \\
Institute of Applied Physics and Computational Mathematics, P.O.
Box 8009 (28), 100088 Beijing, China}

\author{Jing-Ling Chen}

\affiliation{Department of Physics, Faculty of Science, National
University of Singapore}
\title{Geometric Phases for Mixed States during Cyclic Evolutions}

\begin{abstract}
The geometric phases of cyclic evolutions for mixed states are discussed in
the framework of unitary evolution. A canonical one-form is defined whose
line integral gives the geometric phase which is gauge invariant. It reduces
to the Aharonov and Anandan phase in the pure state case. Our definition is
consistent with the phase shift in the proposed experiment [Phys. Rev. Lett.
\textbf{85}, 2845 (2000)] for a cyclic evolution if the unitary
transformation satisfies the parallel transport condition. A comprehensive
geometric interpretation is also given. It shows that the geometric phases
for mixed states share the same geometric sense with the pure states.
\end{abstract}


\maketitle

When a pure quantal state undergoes a cyclic evolution the system
returns to its original state but may acquire a nontrivial phase
factor of purely geometric origin. This was first discovered by
Berry \cite{berry} in the adiabatic context, and generalized to
non-Abelian by Wilczek and Zee \cite {nona}. A nice interpretation
was given by Simon \cite{simon} in terms of a natural Hermitian
connection, as the parallel transport holonomy in a Hermitian line
bundle. Extension to the nonadiabatic cyclic case was given by
Aharonov and Anandan \cite{aa}. Based on Pancharatnam's earlier
work on interference of light \cite{pan}, this concept was
generalized to noncyclic evolutions and nonunitary evolutions
\cite{nonc1,nonc2}. The geometric phases for entangled states have
also been discussed \cite{ent}. Applications of the geometric
phase have been found in molecular dynamics \cite{ek11}, response
function of the many-body system \cite{ek12}, and geometric
quantum computation \cite{ek14}. In all these developments the
geometric phases have been discussed only for pure states.
However, in some applications we are interested in mixed state
cases \cite{ek14,Ekert}.

Uhlmann was probably the first to address the issue of mixed state in the
context of purification, but as a purely mathematical problem \cite{ulman1}.
Recently, Sj\"{o}qvist \textit{et al}. \cite{Ekert} gave a new formalism of
the geometric phase for mixed state in the experiment context of quantum
interferometry under parallel transport condition. It has been pointed out
\cite{com} that the latter geometric phase can be undefined at nodal points
in the parameter space where the interference visibility vanishes. Anyway,
the geometric phases for mixed states proposed in Refs. \cite{ulman1} and
\cite{Ekert} are generically different in the unitary case and match only
under very special conditions such as in terms of pure states \cite{umek}.

In this paper, we give a definition of geometric phase for a cyclic
evolution of mixed quantal state in the dynamical context of quantum system.
The reasons of employing the cyclic evolution are two: (i) the cyclic
evolution of a physical system is of the most interest in physics both
experimentally and theoretically (ii) the phase shift in a cyclic evolution
should be definite. Firstly, we give a straightforward generalization of
Aharonov and Anandan (A-A) phase for the global cyclic evolution where the
total phase is explicit. Though this case seems a trivial extension of (A-A)
phase, it contains the essence of geometric phase of mixed state. Then, we
give the discussion of the general case based on a definition of the total
phase. This geometric phase reduces to the (A-A) phase \cite{aa}, the
standard geometric phase for pure state undergoing a cyclic evolution. We
find that if the evolution satisfies the parallel transport condition the
geometric phase is consistent with the result in Ref.\ \cite{Ekert}.
Moreover, we give the geometric meaning of the geometric phases of mixed
states which share the same sense with the pure states for the first time.

Supposing a quantum system with the Hamiltonian $H(t),$ the density operator
$\rho (t)$ of this system will undergo the following evolution
\begin{equation}
\rho (t)=U(t)\rho (0)U^{+}(t),  \label{evo}
\end{equation}
where $U(t)=\mathbf{T}e^{-i\int_{0}^{t}H(t^{\prime })dt^{\prime }/\hbar }$
is a unitary transformation, here $\mathbf{T}$ is the chronological
operator. If $U(\tau )$ and $\rho (0)$ are commutative: $[U(\tau ),\rho
(0)]=0,$ i.e., $\rho (\tau )=\rho (0),$ we say this state undergoes a cyclic
evolution with period $\tau $. Furthermore, if $U(\tau )=e^{i\phi }I$, this
evolution can be called the global cyclic evolution since for any $\rho (0)$
we have $\rho (\tau )=\rho (0).$

At first, we study the case of the global\ cyclic evolution. Now define $%
\widetilde{U}(t)=e^{-i\phi (t)}U(t)$ such that $\widetilde{U}(\tau )=I.$ We
define the geometric phase for such state during the global cyclic evolution
as
\begin{equation}
\phi _{g}=i\int_{0}^{\tau }Tr\left[ \rho (0)\widetilde{U}^{+}(t)\frac{d%
\widetilde{U}(t)}{dt}\right] dt.  \label{ge2}
\end{equation}
Using the transformation between $\widetilde{U}(t)$ and $U(t),$ one can have
\begin{equation}
\phi _{g}=\phi -\phi _{d},  \label{all}
\end{equation}
and
\begin{eqnarray}
\phi _{d} &=&-i\int_{0}^{\tau }dtTr\left[ \rho (0)U^{+}(t)\frac{dU(t)}{dt}%
\right]  \label{gd} \\
&=&-\frac{1}{\hbar }\int_{0}^{\tau }dtTr\left[ \rho (t)H(t)\right] .
\nonumber
\end{eqnarray}
Obviously $\phi _{d}$ is just the dynamical phase during the cyclic
evolution.

We can prove that if $\rho (t)$ is the density operator of a pure
state the geometric phase defined by Eq. (\ref{ge2}) is just the
Aharonov and Anandan phase \cite{aa}. Assuming $\rho (0)=\left|
\psi (0)\right\rangle \left\langle \psi (0)\right| ,$ $\left| \psi
(t)\right\rangle =U(t)\left| \psi (0)\right\rangle $. Let $\left|
\varphi (t)\right\rangle =e^{-i\phi (t)}\left| \psi
(t)\right\rangle ,$ then we have $\left| \varphi (t)\right\rangle
=\widetilde{U}(t)\left| \varphi (0)\right\rangle .$ So, the Eq.
(\ref{ge2}) can be written as $\phi _{g}=i\int_{0}^{\tau
}\left\langle \varphi (t)\right| \left. \dot \varphi
(t)\right\rangle dt$ which is just the result of Ref. \cite{aa}.

An initial state can always be diagonalized, namely,
\begin{equation}
\rho (0)=\sum\limits_{k}w_{k}\left| k\right\rangle \left\langle k\right| ,
\label{in}
\end{equation}
where $\left| k\right\rangle $ are bases for the system and $w_{k}$ are
classical probability to find a member of the ensemble in the corresponding
state. For the global cyclic evolution we have $U(\tau )\left|
k\right\rangle =e^{i\phi }$ $\left| k\right\rangle ,$ then the A-A phase of $%
\left| k\right\rangle $ is $\phi _{g}^{k}=\phi -\phi _{d}^{k}$, where $\phi
_{d}^{k}=$ $-\frac{1}{\hbar }\int_{0}^{\tau }dtTr\left[ \left|
k\right\rangle \left\langle k\right| U^{+}(t)\dot{U}(t)\right] .$
Substituting (\ref{in}) into (\ref{gd}), and from (\ref{all}) we obtain
\begin{equation}
\phi _{g}=\phi -\sum\limits_{k}w_{k}\phi _{d}^{k}=\sum\limits_{k}w_{k}\phi
_{g}^{k}.  \label{sumg}
\end{equation}
So, for the global cyclic evolution the geometric phases of mixed states
have explicit meanings: the geometric phases of a mixed state is the
weighted average of the geometric phases of the constitute pure states.\

\textit{Example I}. --Suppose that a qubit (a spin-$\frac{1}{2}$ particle)
with a magnetic moment is in a homogenous magnetic field $\mathbf{B}$ along
the $z$ axis. Then the Hamiltonian in the rest frame is $H=-\mu B\sigma
_{z}. $ Suppose the initial state is
\begin{equation}
\rho (0)=\frac{1}{2}[I+r(\sin \theta \sigma _{x}+\cos \theta \sigma _{z})],
\label{ss}
\end{equation}
where $r$ is a constant and $0\leq r\leq 1$. So we have $\rho (t)=U(t)\rho
(0)U^{+}(t)$ with
\begin{equation}
U(t)=\exp (i\mu Bt\sigma _{z}/\hbar ).  \label{ut}
\end{equation}
This unitary evolution is periodic with period $\tau =\pi \hbar /\mu B$,
i.e., $\rho (\tau )=\rho (0).$ It is easy to see that $U(\tau )=\exp (i\pi
)I.$ Let $\widetilde{U}(t)=e^{-i\mu Bt/\hbar }U(t),$ then $\widetilde{U}%
(\tau )=U(0).$ From Eq. (\ref{ge2}), and after some elaboration we can
obtain the geometric phase
\begin{equation}
\phi _{g}=\pi (1-r\cos \theta ).  \label{res}
\end{equation}
Obviously, if $r=1,$ we get $\phi _{g}=\pi (1-\cos \theta )$ which is just
the A-A phase \cite{aa}.

On the other hand, we can have two pure states $\rho _{1}(0)=\frac{1}{2}[%
I+(\sin \theta \sigma _{x}+\cos \theta \sigma _{z})]$ and $\rho _{2}(0)=%
\frac{1}{2}[I+(\sin (\pi +\theta )\sigma _{x}+\cos (\pi +\theta )\sigma _{z})%
],$ which can construct a set of orthonormal bases. \ From
(\ref{ss}), we have $\rho (0)=\frac{1+r}{2}\rho
_{1}(0)+\frac{1-r}{2}\rho _{2}(0).$ Obviously, this is a diagonal
representation of the initial state $\rho (0).$ Then from
Eq.(\ref{sumg}), we can obtain $\phi _{g}=\frac{1+r}{2}\pi (1-\cos
\theta )+\frac{1-r}{2}\pi (1-\cos (\pi +\theta ))=\pi (1-r\cos
\theta ).$

The above discussion of the general cyclic evolution seems a trivial
extension of (A-A) phase, but it contains the essence of geometric phase of
mixed state.

For the general case of a cyclic evolution, the density matrix and
transformation satisfy $[U(\tau ),\rho (0)]=0.$ We can not find the total
phase explicitly from this condition. To factor out the total phase, we use
the Pancharatnam's brilliant idea, i.e., the Pancharatnam connection \cite
{pan}. We define the total phase of the mixed state during a cyclic
evolution with the initial state $\rho (0)$ and the unitary transformation $%
U(t)$ as
\begin{equation}
\phi =\arg Tr[\rho (0)U(\tau )].  \label{tp}
\end{equation}
Let $\widetilde{U}(t)=e^{-i\phi (t)}U(t)$ such that $\phi (\tau )=\phi .$
Based on this definition, the geometric phase of the cyclic evolution can be
also defined by Eq. (\ref{ge2}). Obviously, the geometric phase for a cyclic
evolution takes the same form as in Eq. (\ref{all}). Indeed Eq. (\ref{ge2})
defines a canonical one-form in the parameter space:
\begin{equation}
\beta =iTr\left[ \rho (0)\widetilde{U}^{+}(t)d\widetilde{U}(t)\right] .
\label{g1}
\end{equation}
It is not difficult to prove that $\beta $ is a real number. The geometric
phase can be obtained by its line integral, i.e.,
\begin{equation}
\phi _{g}=\oint \beta .  \label{gee2}
\end{equation}
The equivalent of the above formula for the pure states case is well-known
\cite{pg,nonc2}.

The geometric phase defined above is manifestly gauge invariant: it does not
depend on the dynamics, but it depends only on the geometry of the close
unitary path given by the unitary transformation $U(t).$ Assuming $\alpha $
is a dynamic parameter of this system, the nature of the cyclic evolution
requires $\alpha (0)=\alpha (\tau )$ \cite{berry}$.$ Under the
transformation $\widetilde{U}(t)^{\prime }=e^{i\delta (\alpha )}\widetilde{U}%
(t),$ we can have $\beta ^{\prime }=\beta -d\delta .$ It is easy to prove
that $\oint \beta ^{\prime }=\oint \beta $ since $\delta (a(\tau ))=\delta
(a(0)).$ Indeed the quantity $\beta =iTr[\rho (0)\widetilde{U}^{+}(t)d%
\widetilde{U}(t)]$ can be regarded as a gauge potential on the space of
density operators pertaining to the system.

\textit{Example II.} --Consider a spin-$\frac{1}{2}$\textit{\ }particle is
initially in the state
\begin{equation}
\rho (0)=\frac{1}{2}(I+r\sigma _{z}),\;  \label{int}
\end{equation}
where $r\leq 1$ is a constant. Suppose this particle is in a magnetic field $%
\mathbf{B}(t)$ with
\begin{equation}
\mathbf{B}(t)=\left\{
\begin{array}{cc}
-(\omega \hbar /\mu )\widehat{e}_{y}, & 0\leq t\leq t_{1} \\
-(\omega \hbar /\mu )\widehat{e}_{z}, & t_{1}\leq t\leq t_{2} \\
-(\omega \hbar /\mu )(\sin \varphi \widehat{e}_{x}-\cos \varphi \widehat{e}%
_{y}), & t_{2}\leq t\leq \tau
\end{array}
\right. ,  \label{field}
\end{equation}
in which $\omega $ is a constant and $t_{1}=\frac{\theta }{2\omega },$ $%
t_{2}=$ $\frac{\theta +\varphi }{2\omega }$ and $\tau =\frac{2\theta
+\varphi }{2\omega }.$ So, the unitary transformation is
\begin{equation}
U(t)=\left\{
\begin{array}{cc}
e^{-i\omega t\sigma _{y}}, & 0\leq t\leq t_{1} \\
e^{-i\omega (t-t_{1})\sigma _{z}}e^{-i\frac{\theta }{2}\sigma _{y}}, &
t_{1}\leq t\leq t_{2} \\
\begin{array}{c}
e^{-i\omega (t-t_{2})(\sin \varphi \sigma _{x}-\cos \varphi \sigma _{y})} \\
\times e^{-i\frac{\varphi }{2}\sigma _{z}}e^{-i\frac{\theta }{2}\sigma _{y}},
\end{array}
& t_{2}\leq t\leq \tau
\end{array}
\right. .  \label{tut}
\end{equation}
Then, $U(\tau )=e^{-i\frac{\theta }{2}(\sin \varphi \sigma _{x}-\cos \varphi
\sigma _{y})}e^{-i\frac{\varphi }{2}\sigma _{z}}e^{-i\frac{\theta }{2}\sigma
_{y}}.$ We can prove that $U(\tau )\rho (0)U^{+}(\tau )=\rho (0).$ Under
this unitary transformation, the state undergos a cyclic evolution with a
closed path of the corresponding Bloch vector as showing in Fig.1, where the
vectors at points $B$ and $C$ are $\mathbf{r}_{B}=(r\sin \theta ,0,r\cos
\theta )$ and $\mathbf{r}_{C}=(r\sin \theta \cos \varphi ,r\sin \theta \sin
\varphi ,r\cos \theta )$ respectively. From Eq. (\ref{tp}) and (\ref{gd}),
we have the total phase of this cyclic evolution: $\phi =-\arctan \left[
r\tan \frac{\varphi }{2}\right] ,$ and the dynamic phase: $\phi _{d}=-\frac{%
\varphi }{2}r\cos \theta .$ Then, the geometric phase of this cyclic
evolution is
\begin{equation}
\phi _{g}=-\arctan \left[ r\tan \frac{\varphi }{2}\right] +\frac{\varphi }{2}%
r\cos \theta .  \label{rre}
\end{equation}
We know that if $\theta =\frac{\pi }{2}$ the closed path on the
Bloch sphere is geodesic \cite{ek21}, i.e., the dynamical phase is
zero. At this time the geometric phase is $-\arctan \left[ r\tan
\frac{\varphi }{2}\right] ,$ which has also been pointed out in
Ref. \cite{Ekert}  and verified by the recent experimental
observations \cite{exp}.


\begin{figure}[!htb]
\begin{center}
\rotatebox{-90}{\resizebox *{8.0cm}{7.0cm} {\includegraphics
{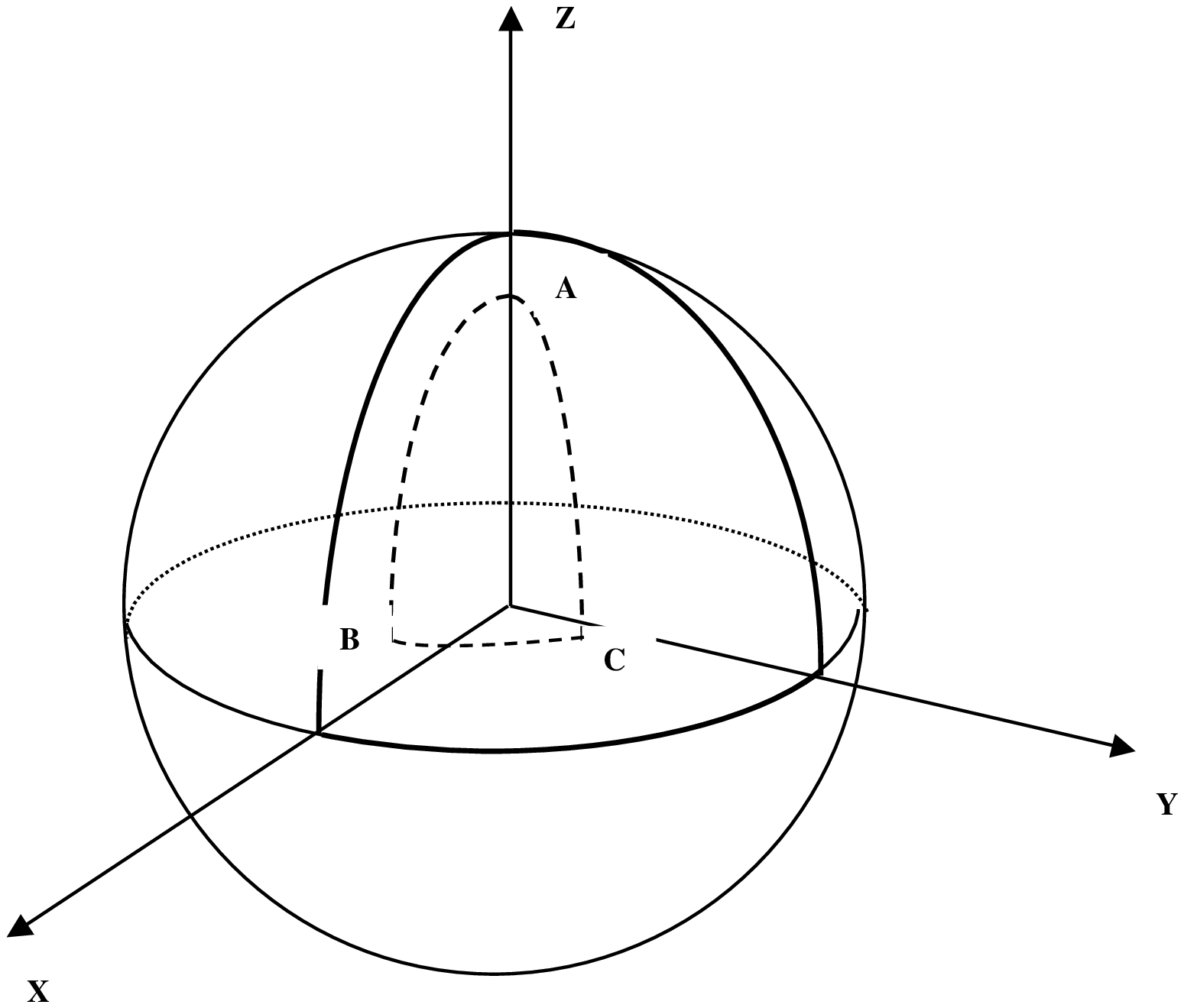}}}
\end{center}
\caption{Path with cyclic evolution of the Bloch vector. The solid line
represents a geodesic path of the unitary pure state case and defines a
spherical triangle enclosing the solid angle $\protect\pi /2.$ The dashed
line represents the closed path of the mixed state with the unitary
evolution defined by (\ref{tut}), where the initial state is given in Eq. (%
\ref{int}).}
\label{fig1}
\end{figure}

If $\varphi =2\pi ,$ i.e., the points $B$ and $C$ in Fig. 1 are the same.
Obviously, the geometric phase for such a path should be equal to the case
in \textit{Example I} except for the direction of magnetic field reversed.
From (\ref{rre}), we have $\phi _{g}=\pi (1+r\cos \theta )$ which is
consistent with Eq. (\ref{res}).

From our definition of the geometric phase, we can get a theorem for a
composite quantum system.

\textit{Theorem:} Let $\rho ^{AB}$ be a density operator of a composite
quantum system of $A$ and $B$. \ If the system evolves under the unitary
transformation $U(t)=I^{A}$ $\otimes U^{B}$ with $U^{B}=\mathbf{T}%
e^{-i\int_{0}^{t}H(t^{\prime })dt^{\prime }/\hbar }$ and $[\rho ^{AB},U(\tau
)]=0$, then the geometric phase of such a composite system equals to one of
the subsystem $B,$ i.e., $\phi _{g}^{AB}=\phi _{g}^{B}.$

The density operator of such system can be expressed by
\begin{eqnarray}
\rho ^{AB} &=&\frac{1}{N_{A}N_{B}}[I^{A}\otimes I^{B}+\mathbf{r}^{A}\cdot
\vec{\lambda}^{A}\otimes I^{B}  \nonumber \\
&&+I^{A}\otimes \mathbf{r}^{B}\cdot \vec{\lambda}^{B}+\beta _{ij}\lambda
_{i}^{A}\otimes \lambda _{j}^{B}],  \label{rab}
\end{eqnarray}
where $N_{A}$ and $N_{B}$ are the orders of density matrices for each
subsystems, $\vec{\lambda}^{A}=(\lambda _{i}^{A};i=1,2,\cdots ,N_{A}^{2}-1)$
and $\vec{\lambda}^{B}=(\lambda _{i}^{A};i=1,2,\cdots ,N_{B}^{2}-1)$ are the
generators of $SU(N_{A})$ and $SU(N_{B})$ respectively, $\mathbf{r}%
^{A}=(r_{i}^{A};i=1,2,\cdots ,N_{A}^{2}-1)$ and $\mathbf{r}%
^{B}=(r_{i}^{B};i=1,2,\cdots ,N_{B}^{2}-1)$ are two vectors, and
$\beta _{ij} $ are $(N_{A}^{2}-1)(N_{B}^{2}-1)$ real numbers.
Because $Tr(\lambda _{i}^{A,B})=0,$ one can have $\phi ^{AB}=\arg
Tr[\rho ^{AB}U(\tau )]=\arg Tr[\rho ^{B}U^{B}(\tau )]$ where $\rho
^{B}$ is reduced density operator for $B,$
i.e., $\rho ^{B}=\frac{1}{N_{B}}[I^{B}+$ $\mathbf{r}^{B}\cdot \vec{\lambda}%
^{B}]$. From (\ref{gd}), we can also obtain $\phi _{d}^{AB}=-i\int
Tr[\rho ^{AB}U^{+}(t)\dot{U}(t)]=-i\int Tr[\rho
^{B}U^{B^{+}}(t)\dot{U}^{B}(t)]$. So the geometric phase for $\rho
^{AB}$ is $\phi _{g}^{AB}=\phi ^{AB}-\phi _{d}^{AB}$ $=\phi
^{B}-\phi _{d}^{B}$, where $\phi ^{B}=\arg Tr[\rho
^{B}U^{B}(\tau )]$ and $\phi _{d}^{B}=-i\int Tr[\rho ^{B}U^{B^{+}}(t)\dot{U}%
^{B}(t)].$ The geometric phase of the subsystem $B$ is $\phi _{g}^{B}$ $%
=\phi ^{B}-\phi _{d}^{B}=\phi _{g}^{AB}.$

We know that for a pure state of an entangled composite quantum system $\rho
^{AB},$ the reduced density operators $\rho ^{A,B}$ are mixed states. So
this theorem may have some important applications, at least to observe the
geometric phase for mixed state, since the geometric phase for a pure state
can be acquired by methods known before.

Recently, Sj\"{o}qvist \textit{et al}. have also proposed geometric phases
for mixed states under the parallel transport condition \cite{Ekert}. If the
unitary transformation $U(t)$ satisfies a parallel condition we can prove
the geometric phase given by Eq. (\ref{ge2}) is consistent with what was
obtained in Ref. \cite{Ekert}. The parallel transport condition for a mixed
state undergoing unitary evolution is
\begin{equation}
Tr[\rho (t)U^{+}(t)\dot{U}(t)]=0.  \label{parall}
\end{equation}
Under this condition, from the definition in Ref. \cite{Ekert} the geometric
phase of the cyclic evolution can be expressed as $\overline{\phi }_{g}=\arg
Tr[\rho (0)U(\tau )].$

Indeed, for the unitary operator $U(t)$ we can always obtain another unitary
operator by a $U(1)$ transformation, $U^{\prime }(t)=e^{i\xi (t)}U(t),$ so
that $Tr[\rho (t)U^{\prime +}(t)\dot{U}^{\prime }(t)]=0.$ From this
condition, it is easy to prove that
\begin{equation}
\dot{\xi}(t)=iTr[\rho (0)U^{+}(t)\dot{U}(t)].  \label{aaa}
\end{equation}
So $\xi (\tau )=-\phi _{d}.$ From (\ref{tp}) and (\ref{all}),
$\phi _{g}=\arg Tr[\rho (0)U(\tau )]+\xi (\tau )=\arg Tr[\rho
(0)U^{\prime }(\tau )].$ Obviously, such a parallel transport
$U^{\prime }(\tau )$ is unique for an initial state $\rho (0)$ and
the unitary operator $U(t)$. So, if $U(t)$ satisfies the parallel
transport condition, the dynamical phase in Eq. (\ref {all})
vanishes identically, so $\overline{\phi }_{g}=\phi _{g},$ i.e.,
under the parallel transport condition our definition of the
geometric phase is consistent with the definition in Ref.
\cite{Ekert}.

However, for an initial state $\rho (0)$ and the unitary operator
$U(t),$ one can construct alternative parallel transport by a
unitary transformation, except for a $U(1)$ transformation, i.e.,
the parallel transport corresponding the initial state and the
unitary operator $U(t)$ is not unique \cite{Ekert,cond}. We must
emphasize that only for the parallel transport obtained by the
$U(1)$ transformation (which is unique), $U^{\prime }(\tau
)=e^{i\xi (t)}U(t)$ determined by Eq. (\ref {aaa}), the geometric
phase can be expressed as $\phi _{g}=\arg Tr[\rho
(0)U^{\prime }(\tau )]$. As an example, we consider the case of \textit{%
Example I}. For the state $\rho (0)$ given in (\ref{ss})$,$ one can have two
pure states $\rho _{\pm }=\frac{1}{2}[I\pm (\sin \theta \sigma _{x}+\cos
\theta \sigma _{z})]$ such that $\rho (0)=\frac{1+r}{2}\rho _{+}+\frac{1-r}{2%
}\rho _{-}.$ For the unitary operator $U(t)=\exp (i\mu Bt\sigma _{z}/\hbar ),
$ by unitary transformation $e^{-i[\mu Bt(\cos \theta )(\sin \theta \sigma
_{x}+\cos \theta \sigma _{z})/\hbar ]},$ we can construct $U^{\prime \prime
}=e^{i(\mu Bt\sigma _{z}/\hbar )}e^{-i[\mu Bt(\cos \theta )(\sin \theta
\sigma _{x}+\cos \theta \sigma _{z})/\hbar ]}$  which satisfies $Tr[\rho
_{\pm }(U^{\prime \prime })^{+}\dot{U}^{\prime \prime }(t)]=0.$ Obviously,
this is a parallel transport of $\rho (t)=U^{\prime \prime }(t)\rho
(0)U^{\prime \prime ^{+}}(t)=U(t)\rho (0)U^{+}(t).$ It is easy to obtain $%
\arg Tr[\rho (0)U^{\prime \prime }(\tau )]=\pi -\arctan [r\tan (\pi \cos
\theta )]$ which does not equal to the geometric phase given by Eq. (\ref
{res}).

In fact, the set of the density operators constructs a base space $\mathcal{M%
}$ (the subspace corresponding to pure states is isomorphic to the
projective Hilbert space)$\mathcal{.}$ The unitary transformations
$U$ just define a line bundle $\mathcal{F}$ on the base space with
the projection map $\Pi $ as $\Pi ^{-1}(\rho )=\{\rho :\rho
\rightarrow c=Tr(U\rho )/|Tr(U\rho )|,\;c$ $\in U(1)\}$. Then the
cyclic evolution discussed in the present paper is just define a
closed curve $\mathit{C}:[0,\tau ]\rightarrow \mathcal{M}$ on the
base space, but on the bundle space the path is not closed with
the final point lifted by a $U(1)$ factor $c(\tau ).$ On the other
hand, using $U(1)$ transformations, we can always obtain a
unitary transformation $U^{\prime }(t)$ determined by Eq. (\ref{aaa}%
) which satisfies the parallel transport condition for a given
closed path $C.$ Then, the geometric phase defined in Eq.
(\ref{all}) can be expressed by the phase factor of the parallel
lift : $\phi _{g}=\arg Tr[\rho (\tau )U^{\prime }(\tau )]$. Under
this consideration the geometric phases of mixed states share the
same geometric sense with the pure states for the cyclic
evolution.

In summary, we give a definition of the geometric phase for a mixed state
during a cyclic evolution. Our definition is consistent with the
prescription proposed in Ref. \cite{Ekert} when the evolution satisfies
parallel transport condition. We first give the geometric meaning of the
geometric phases for mixed states which share the same sense with the pure
states. Although our definition is only for cyclic evolution, the discussion
can be straightforwardly generalized to noncyclic evolution except for a
gauge condition \cite{gauge} (for the pure states see in \cite{nonc2}) which
is automatically satisfied in cyclic evolution.

We thank Professor S.G. Chen and Professor X.G. Zhao for useful
discussions. This project was supported by Fundamental Research
Project (973 Project) of China. LB Fu acknowledges the Alexander
von Humboldt Foundation for supporting.


\begin{thebibliography}{99}
\bibitem{berry}  M.V. Berry, Proc. R. Soc. London A \textbf{392}, 45 (1984).

\bibitem{nona}  F. Wilczek and A. Zee, Phys. Rev. Lett. 52, 2111 (1984).

\bibitem{simon}  B. Simon, Phys. Rev. Lett. \textbf{51}, 2167 (1983).

\bibitem{aa}  Y. Aharonov and J.S. Anandan, Phys. Rev. Lett. \textbf{58},
1593 (1987).

\bibitem{pan}  S. Pancharatnam, Proc. Indian Acad. Sci. A \textbf{44}, 247
(1956).

\bibitem{nonc1}  J. Samuel and R. Pistolesi, Phys. Rev. Lett. \textbf{60},
2339 (1988);

\bibitem{nonc2}  A. K. Pati, Phys. Rev. A \textbf{52}, 2576 (1995); A.K.
Pati, J. Phys. A: Maht. Gen. \textbf{28}, 2087 (1995).

\bibitem{ent}  E. Sj\"{o}qvist, Phys. Rev. A \textbf{62}, 022109 (2000).

\bibitem{ek11}  E. Sj\"{o}qvist and H. Hedstr\"{o}m, Phys. Rev. A \textbf{56}%
, 3417 (1997).

\bibitem{ek12}  S.R. Jain and A. K. Pati, Phys. Rev. Lett. \textbf{80}, 650
(1998); A.K. Pati, Phys. Rev. A \textbf{60}, 121 (1999).

\bibitem{ek14}  J.A. Jones, V. Vedral, A. Ekert, and G. Castagnoli, Nature
(London) \textbf{403}, 869 (1999).

\bibitem{Ekert}  E. Sj\"{o}qvist, A.K. Pati, A. Ekert, J.S. Anandan, M.
Ericsson, D.K.L. Oi, and V. Vedral, Phys. Rev. Lett. \textbf{85}, 2845
(2000).

\bibitem{ulman1}  A. Uhmann, Rep. Math. Phys. \textbf{24}, 229 (1986); A.
Uhlmann, Lett. Math. Phys. \textbf{21} 229 (1991).

\bibitem{com}  R. Bhandari, Phys. Rev. Lett. \textbf{89}, 268901 (2002); J.
S. Anandan, E. Sj\"{o}qvist, A.K. Pati, A. Ekert, M. Ericsson, D.K.L. Oi,
and V. Vedral, Phys. Rev. Lett. \textbf{89}, 268902 (2002).

\bibitem{umek}  M. Ericsson, A.K. Pati, E. Sj\"{o}qvist, J. Br\"{a}nnlund,
and D.K.L. Oi, e-print quant-ph/0206063; J. Tidstr\"{o}m and Sj\"{o}qvist,
e-print quant-ph/0211187.


\bibitem{pg}  D.N. Page, Phys. Rev. A \textbf{36}, 3479 (1987).

\bibitem{ek21}  J.S. Ananda, Phys. Lett. A \textbf{129}, 201 (1988).

\bibitem{exp} J. Du, P. Zou, M. Shi, L.C. Kwek, J.-W, Pan, C.H.
Oh, A. Ekert, D.K.L. Oi, and M. Ericsson, Phys. Rev. Lett.
\textbf{91}, 100403 (2003).

\bibitem{cond}  A.G. Wagh and V.C. Rakhecha, Phys. Lett. A \textbf{197}, 107
(1995).

\bibitem{gauge}  Under a $U(1)$ transformation $\widetilde{U}(t)^{\prime
}=e^{i\delta (\alpha )}\widetilde{U}(t),$ we can have $\beta ^{\prime }=iTr%
\left[ \rho (0)\widetilde{U}^{\prime +}(t)d\widetilde{U}^{\prime }(t)\right]
=\beta -d\delta .$ So the gauge function should satisfy: $\delta (\tau
)=\delta (0)+2n\pi $ where $n$ is a integer number. The phase angle will be
invariant mod $2\pi $.
\end{thebibliography}
\end{document}